DEC-TR-592

# Characteristics of Destination Address Locality in Computer Networks: A Comparison of Caching Schemes

Raj Jain

Digital Equipment Corporation
550 King St. (LKG1-2/A19)
Littleton, MA 01460

Network Address: Jain%Erlang.DEC@DECWRL.DEC.COM

# Characteristics of Destination Address Locality in Computer Networks: A Comparison of Caching Schemes

Raj Jain
Distributed Systems Architecture & Performance
Digital Equipment Corp.
550 King St. (LKG 1-2/A19)
Littleton, MA 01460
ARPAnet: Jain%Erlang.DEC@DECWRL.DEC.COM

**Abstract**

The size of computer networks, along with their bandwidths, is growing exponentia
high-speed networks, it is necessary to be able to forward packets in a few mic
forwarding operation consists of searching through a large address database. Th
design of adapters, bridges, routers, gateways, and name servers.

Caching can reduce the lookup time if there is a locality in the address referenc
reference trace measured on an extended local area network, we attempt to see if
have a significant locality.

We compared the performance of MIN, LRU, FIFO, and random cache replacement algor
the interactive (terminal) traffic in our sample had quite different locality behavio
traffic. The interactive traffic did not follow the LRU stack model while the noninte
are shown of the environments in which caching can help as well as those in which
cache size is large.

## 1 INTRODUCTION

The fact that page references by computer programs exhibit locality behavior is now well established and designing computer systems without virtual memory and memory caches is practically inconceivable [20, 27]. In the 1970s there were a large number of studies of program behavior [16, 26] that helped design several good page replacement algorithms and caching strategies. In the 1980s, with the increasing trend towards distributed computing, the caching of files (located remotely) and the study of file reference behavior became an interesting topic [5, 6, 13, 14, 17, 21, 23, 28, 29].

Recently, we discovered that the frames on computer networks also exhibit locality behavior [9]. The understanding of this behavior will help us design the large networks of the 1990s in an efficient manner.

The trend toward networks becoming larger and faster, and addresses also increasing in s
pelled a need to understand and exploit the
if one exists. DECnet Phase IV currently a
to 64,000 nodes and DEC's internal networki
EasyNet [18], has more than 30,000 nodes.
large networks obviously need more efficient
lookups. The size of the addresses themsel
growing. HDLC, a commonly used datalink p
col standard, was designed with 8-bit addr
IEEE 802 LAN protocols support 48-bit add
and the ISO/OSI network layer requires 160
octets) addresses. This increased length
has also necessitated a need to find efficient
look up addresses. Finally, as networks a
ing faster, network routers, which previou
a few hundred frames per second, are now ex
to handle 8000 to 16,000 frames per secon
fast handling requires squeezing every cy
the frame forwarding code.

The realization that the frame destinations cality behavior makes caching a possi[illegible] for efficiently supporting large networks. By caching the destinations recently seen, the intermediate nodes can avoid looking through large tables of nodes with a high probability. The address space need not be hierarchical, the caching works with flat as well as hierarchical address. Caching is transparent in that no protocol changes are generally required to accommodate caching and noncaching implementations in the same network.

The cost of memory chips has been falling rapidly; however, their access times have not decreased as fast. As a result, although the cost of the memory to hold these large address databases may not be a significant consideration (as was the case for development of virtual memory)i, but the access time of the address database is the major reason for our need to find efficient ways to look up addresses. Caching allows such decisions to be made correctly within the specified time limit with a high probability. Since incorrect decisions may result in frames being retransmitted, the cache should be designed so that a very low miss probability will result, typically less than 0.1%. This should be contrasted with page replacement algorithms, where miss probability of 10% may be considered acceptable.

In this paper, we are concerned with the problem of address recognition in bridges. However, there are a number of other applications in computer networks where caching can help avoid searching through a number of entries. For example, datalink adapters can use caching to search through the list of multicast addresses. The network adapter board [11] uses caching to help decode the received frame header. Routers and gateways can cache forwarding databases. Also, name servers and their clients can use caching to improve the efficiency of name lookup. Although, the conclusions of our reference trace are not applicable to these other applications, our methodology, when applied to traces of these applications, can be used to find the appropriate caching strategy.

The organization of this paper is as follows. First, we describe the environment in which the address trace was measured. Second, we explain various locality concepts and analyze the applicability of different locality models. We then compare the performance of various cache replacement algorithms.

## 2 Measured Environment

In order to compare various caching strat[illegible] used a trace of destination addresses obs[illegible] an extended local area network in use at [illegible] King Street, Littleton facility. The netw[illegible] of several Ethernet LANs interconnected vi[illegible] The network is a part of Digital's company-[illegible]work called EasyNet [18], which has more tha[illegible] nodes. The building itself has approxima[illegible] nodes on several Ethernet LANs interconne[illegible] bridges. There are 30 Level-1 routers, s[illegible] routers, and approximately 80 bridges in t[illegible]ing. A promiscuous monitor attached to one Ethernet LAN produced a time-stamped ref[illegible] string of approximately 2 million frames. [illegible] analyses we subdivided the trace into 11 [illegible] of approximately 200,000 frames each. Th[illegible]acteristics of these subtraces along with [illegible] complete trace are listed in Table 1.

Table 1: Trace Characteristics

| Subtrace # | Frames | Addresses Total | Addresses Dest. | Hours |
|---|---|---|---|---|
| 1 | 200000 | 460 | 244 | 0.12 |
| 2 | 200000 | 450 | 208 | 0.12 |
| 3 | 200000 | 449 | 210 | 0.11 |
| 4 | 200000 | 437 | 210 | 0.11 |
| 5 | 200000 | 435 | 203 | 0.11 |
| 6 | 200000 | 436 | 204 | 0.10 |
| 7 | 200000 | 444 | 201 | 0.11 |
| 8 | 200000 | 433 | 205 | 0.10 |
| 9 | 200000 | 424 | 210 | 0.09 |
| 10 | 200000 | 431 | 207 | 0.10 |
| 11 | 46000 | 379 | 186 | 0.02 |
| Total | 2046000 | 495 | 296 | 1.09 |

The total column includes addresses in dest[illegible] well as source fields of the frame. This numb[illegible]proximately equal to the number of station[illegible] extended LAN, since all stations periodica[illegible]cast a 'hello' message to indicate their pr[illegible] the network. Not all addresses appear in th[illegible]tion address field since only a fraction of i[illegible] addressed (unicast) frames pass through t[illegible]tored LAN. For example, in subtrace 1, the[illegible] 460 distinct addresses; of these, only 244 a[illegible] the destination address fields. Due to bri[illegible]ing, only those frames whose desinations ha[illegible] path through the monitored segment are see[illegible] segment. The hour column gives the duration of the subtrace in hours. As shown in the table, the complete trace was a result of approximately one hour of monitoring.

There are several advantages and disadvantages of using a trace. A trace is more credible than references generated randomly using a distribution. On the other hand, traces taken on one system may not be representative of the workload on another system. We hope that others will find the methodology presented here useful and will apply it to traces of environments relevant to their applications.

## 3 Locality: Concepts

In this section we review some of the well-known concepts about locality. These concepts were developed during studies of page reference patterns, but apply equally well to file reference or destination reference patterns. In the following discussion, the term *address* refers to page, file, or the destination node encountered.

The locality of a reference pattern may be temporal or spatial. Temporal locality implies a high probability of reuse. For example, the reference string {3, 3, 3, 3, 3, ...} has a high temporal locality, since the address 3 is used repeatedly once it is referenced. Spatial locality implies a high probability of reference to *neighboring* addresses. For example, the string {1, 2, 3, 4, 5, ...} has a high spatial locality since after a reference to address $k$, the probability of reference to $k+1$ is very high. While the definition of *neighboring* addresses is somewhat clear for page and file addresses, it is not so clear for networks. Spatial locality, if present, is useful in designing prefetching algorithms since the information likely to be used in the near future is fetched before its first reference, thereby, avoiding a *cache miss*. Page reference patterns exhibit both temporal as well as spatial locality.

The terms *persistence* and *concentration* have also been used to characterize locality behavior [2]. Persistence refers to the tendency to repeat the use of a single address. This is, therefore, similar to temporal locality. Concentration, on the other hand, refers to the tendency of the references to be limited (concentrated) to a small subset of the whole address space. A reference string with high concentration is good in that a small cache would produce large performance gains. Persistence can be measured by counting consecutive references to the same address, while concentration can be measured by computing the fra address space used for a large fraction of t string. For example, in a reference string persistence, the probability of the same ad referenced consecutively may be 60%, for Similarly, in a string with high concentra of the references may be to 1% of the addres Bunt and Murphy [2] have done extensive stu persistence and concentration in memory an erence strings.

Virtual memory is one of the first applica locality concepts in computer systems des pages actively being used are kept in the (cache) memory. The key differences between memory, file caching, and destination addres are summarized in Table 2. In virtual memo tems, a very large cache (physical memory) g ter performance, but is too expensive. In n systems, large local caching not only requ local memory, but also results in a large am information being transported over the netwo In this case, there is an optimal cache size the caching does not pay. This is true for d address caching too. If the cache is too big time is large and caching is not useful. T caches may result in too many page faults i memory systems or too many network accesses mote file systems. In either case, the syst wait while the information is being fetche increased response time. This is also tru nation address caching. A long delay in add up may result in the source retransmitting The cache miss rate has to be kept low. Acc miss rates range from 0.1% to 10% depending the ratio of lookup time with and without t A larger ratio would increase the probabil transmissions and would need a smaller mis

## 4 Models of Reference Behavior

A number of models have been developed for p erence behavior. These well known models ar dependent reference model (IRM), the least used (LRU) stack model, and the working set model. In the following subsections, we des models and see their applicability to our a erence trace.

Table 2: Locality in Page vs File vs Node References

| | Page | File | Node |
|---|---|---|---|
| Year needed | 1970 | 1980 | 1990 |
| Why needed | Large programs | Remote files | Large networks |
| Why not infinite cache | Memory cost | Memory cost & comm. overhead | Access time |
| Cost of miss | Page fault | Network access | Packet lost or delayed |
| Effect of a high miss rate | Thrashing | | Instability |
| Good miss rate | 10% | 1% | 0.1% to 10% |

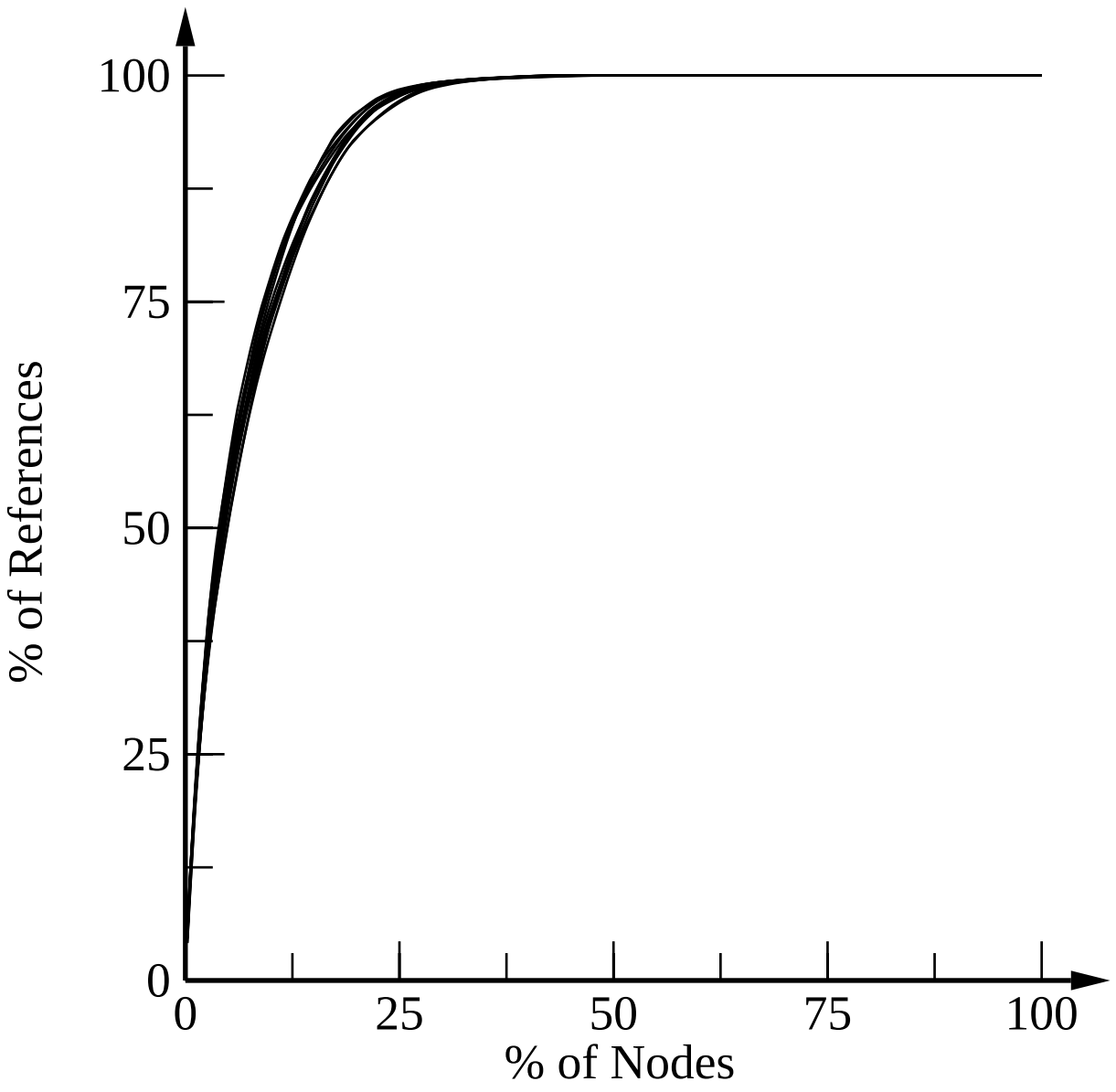

Figure 1: Percentage of frames vs percenta tinations.

### 4.1 Independent Reference Model

The independent reference model assumes, as the name implies, that the references are independent [19]. Knowing that the last reference was to address $k$ does not give any information about the next address to be referenced. In other words, this model assumes that the reference strings do not have any temporal or spatial locality. The probability of reference to address $i$ is $p_i$, and $p_i$'s need not be equal. In a more restricted IRM, called Uniform-IRM, the probability $p_i$'s are assumed to be all equal. This is equivalent to assuming that there is no concentration of references.

Figure 1 show the cumulative frequency of reference as a function of fraction of distinct addresses seen in the trace. Notice that the destination reference probability is nonuniform. For uniform probability, the curve would have been a straight line between (0%, 0%), and (100%, 100%). The median and 90-percentile points on the curves are listed in Table 3.

Notice that 50% of the frames are destined to 4% of the destinations and that 90% of the frames are destined to 17% of the destinations. Thus, destination references exhibit a strong *concentration*. This is a good news since it implies that if we cache highly probable destinations, we may get high hit rates with small caches.

Table 3: Cumulative Percentage of Refere

| Subtrace | Median | 90-Perc |
|---|---|---|
| 1 | 4.1 | 15.7 |
| 2 | 4.2 | 15.6 |
| 3 | 4.7 | 16.9 |
| 4 | 4.8 | 16.9 |
| 5 | 5.1 | 17.7 |
| 6 | 5.0 | 17.0 |
| 7 | 4.5 | 15.5 |
| 8 | 4.4 | 16.4 |
| 9 | 4.0 | 16.0 |
| 10 | 4.6 | 17.2 |
| 11 | 4.7 | 17.2 |
| Total | 4.4 | 17.8 |

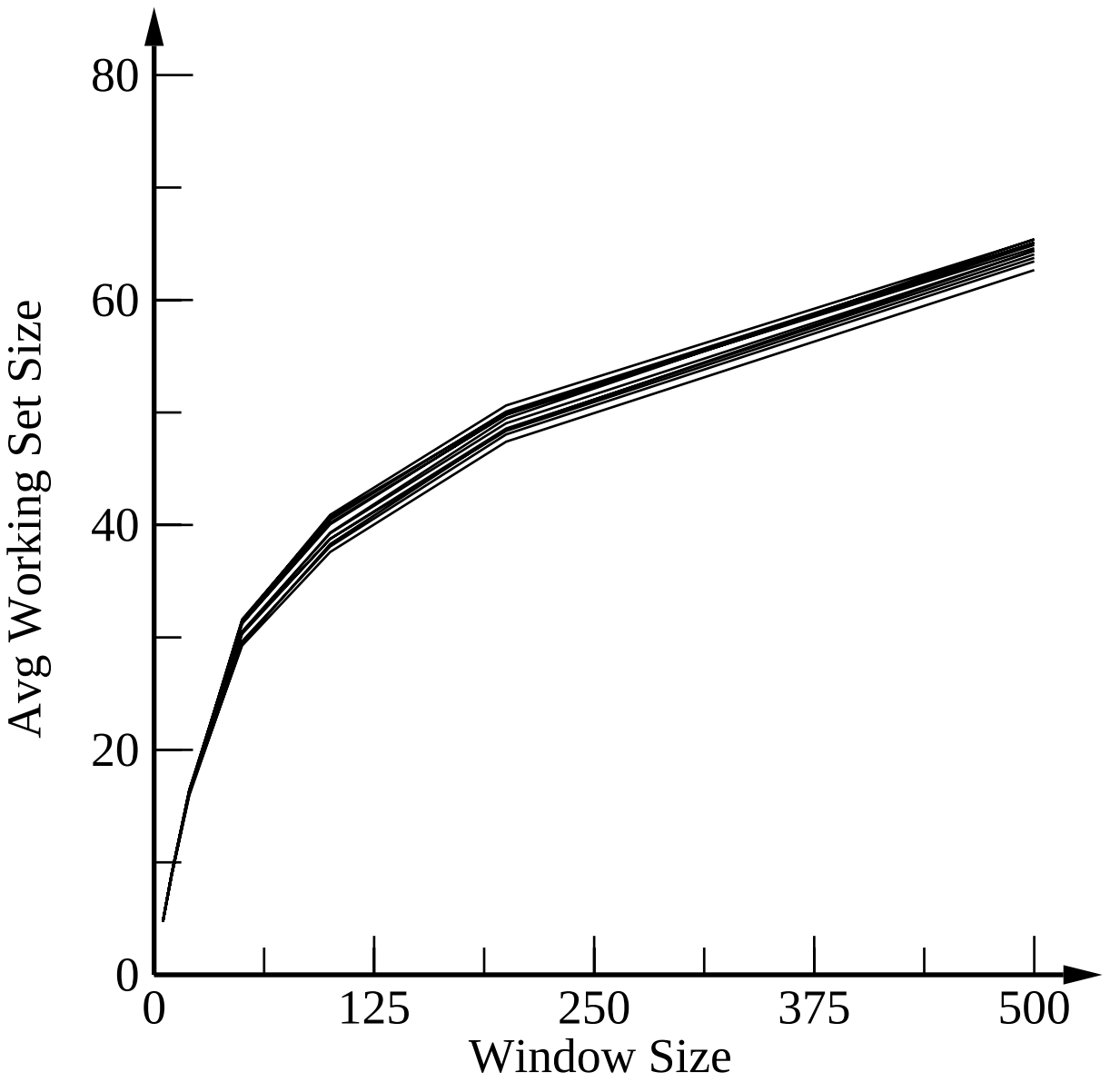

Figure 2: Working set size.

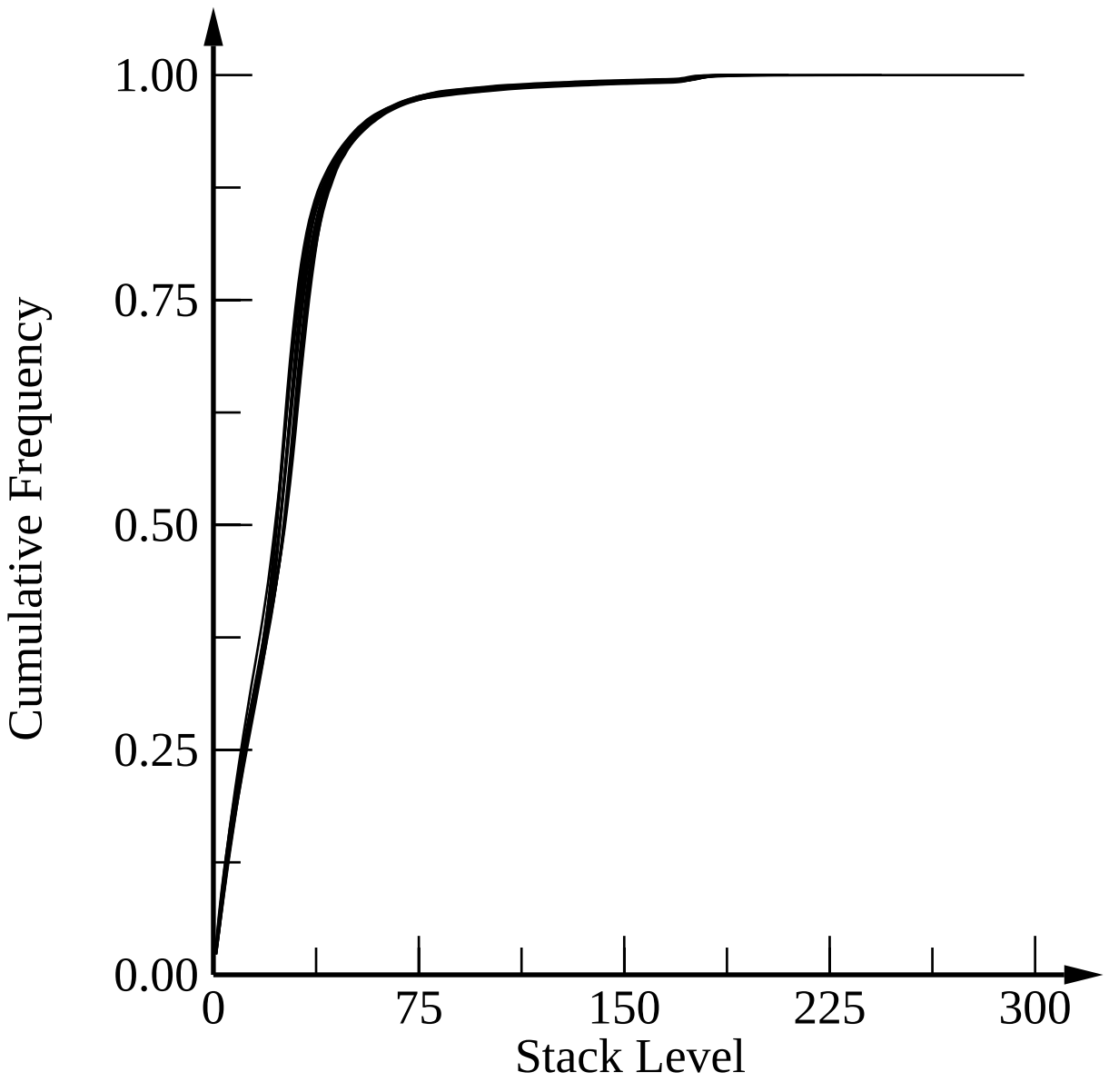

Figure 3: Stack distance cumulative proba
tribution function.

Another distinct feature of Figure 1 is that all subtraces have almost identical behavior. Since these traces consist of traffic during different time intervals on the same network, the observed behavior does not seem to be a reflection of a short-term activity.

## 5 Working Set Model

The working set model [3] assumes that the addresses referenced in the last $W$ references are highly likely to be rereferenced. The interval $W$ is called the **working set window size**, and the number of distinct references in the interval is called the **working set size**. High temporal locality is reflected by a small working set size.

Figure 2 shows the average working set sizes for several different window sizes. The data shows that the destination reference pattern has a high temporal locality. For example, 65 distinct destinations were referenced on the average in successive working set windows of 500 references. In the absence of temporal locality, this number should have been close to 500.

Also, notice that the temporal locality does not exist for small working set window sizes (of up to 50). For example, the average working set size for a window of 10 references is 9.

## 6 LRU Stack Model

The LRU stack model assumes that the proba
of reference to an address is a decreasing
time since it was last referenced. If the a
arranged in a stack so that the address ref
always taken out of its current position i
and pushed to the top of the stack, the proba
of $i$th stack position (counting from the top
the bottom of the stack) being referenced is
ing function of $i$. For a reference string w
temporal locality, the probability $p_1$ of the stack top
being referenced again would be high. Thi
has been analyzed extensively in literatur
with [22].

The cumulative frequency of reference up t
different stack levels is shown in Figure 3
that:

1. The stack top (level 1) reference fre
only 2% to 3%. This is different from the
measured at M.I.T. [4, 9] where 30% of t
erences were found at the stack top and t
two levels had a cumulative reference fr
of 60%.
2. We see that the top 100 stack positions
the total possible stack positions) ac

98% of the frames. This is much lower than corresponding figures seen for page reference and file reference strings [2].

The first observation above is further substantiated by a study of consecutive references. Table 4 shows the observed frequency of a destination being referenced in *n* successive frames for various values of *n*. Notice that the frequencies are rather small.

Table 4: Frequency of Consecutive References

| Subtrace | Number of Consecutive References 1 | 2 | 3 | 4 | Longest |
|---|---|---|---|---|---|
| 1 | 0.946 | 0.024 | 0.001 | 0.000 | 10 |
| 2 | 0.948 | 0.023 | 0.001 | 0.000 | 8 |
| 3 | 0.955 | 0.021 | 0.001 | 0.000 | 8 |
| 4 | 0.940 | 0.026 | 0.002 | 0.001 | 9 |
| 5 | 0.947 | 0.023 | 0.001 | 0.000 | 14 |
| 6 | 0.955 | 0.021 | 0.001 | 0.000 | 10 |
| 7 | 0.948 | 0.023 | 0.001 | 0.000 | 9 |
| 8 | 0.947 | 0.022 | 0.002 | 0.000 | 8 |
| 9 | 0.936 | 0.025 | 0.003 | 0.000 | 9 |
| 10 | 0.946 | 0.024 | 0.002 | 0.000 | 9 |
| 11 | 0.957 | 0.020 | 0.001 | 0.000 | 5 |
| Total | 0.947 | 0.023 | 0.001 | 0.000 | 14 |

## 7 Cache Replacement Algorithms

More important than the theoretical question of which locality model applies best to the destination references is the practical question of which replacement algorithm is best for caching such addresses. To answer this latter question, we compared different cache replacement algorithms. The traditional metric for performance of a cache is the number of *faults* or *misses*. A fault or miss is said to occur when an address is not found in the cache. On a cache miss, one of the entries in the cache must be replaced to bring in the missed entry. Several replacement algorithms can be found in the literature on processor design and virtual memory. We chose four popular algorithms for comparison: least recently used (LRU), first in first out (FIFO), random (RAND), and a theoretically optimal algorithm called MIN [1]. Given a reference trace and a fixed-size cache, it has been proven that the MIN algorithm would cause less faults than any other algorithm. MIN chooses the address that will be referenced farthest in future. It, therefore, requires looking ahead in the reference string.

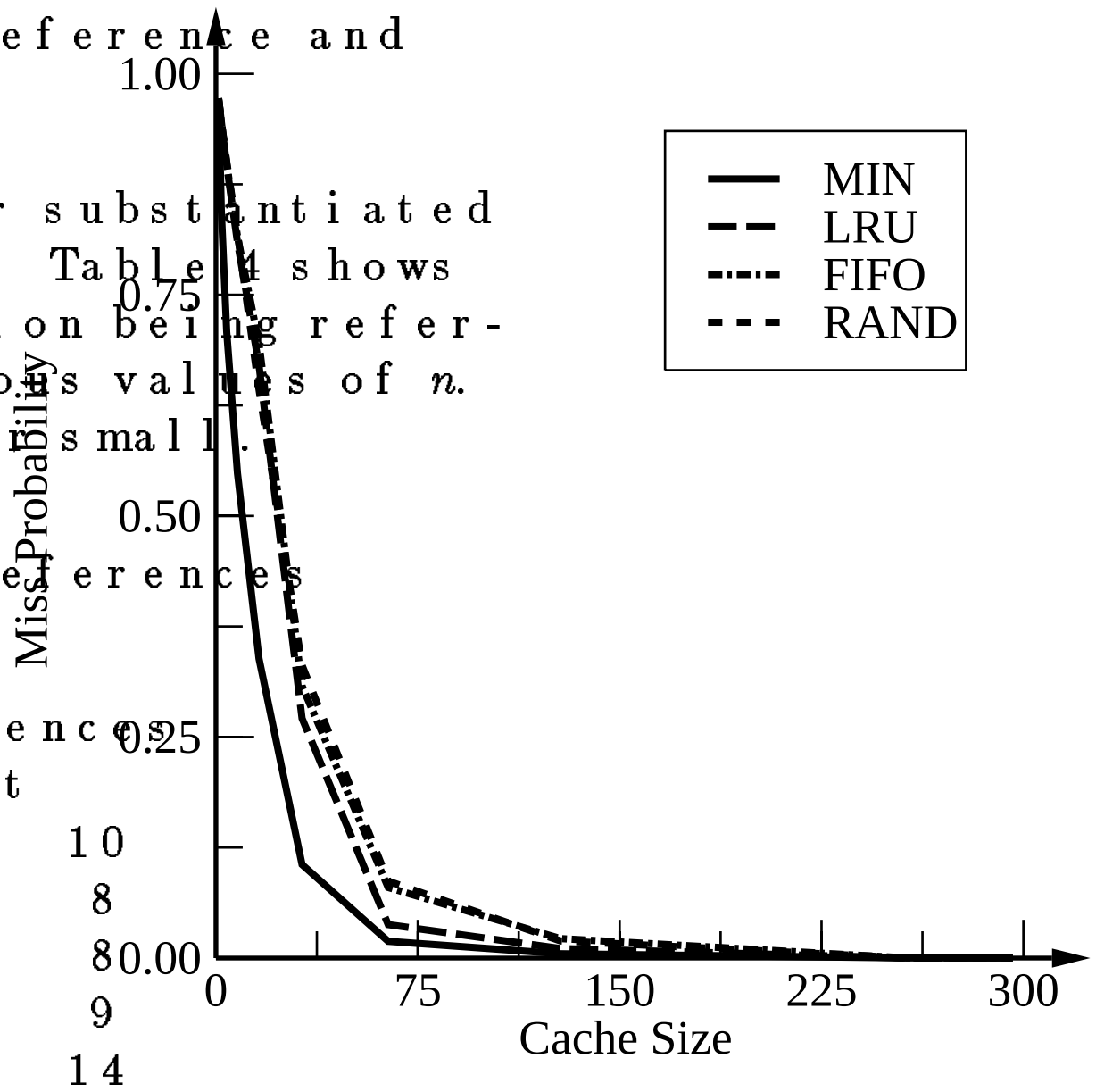

Figure 4: Cache miss probability for vari
replacement algorithms.

Obviously, it cannot be implemented in a r tem. Nonetheless, it provides a measure of particular algorithm is from the theoretic

We used the following three metrics to comp replacement algorithms:

1. Miss probability
2. Interfault distance
3. Normalized search time

We have defined these metrics and the resul presented in the following subsections.

### 7.1 Miss Probability

The miss probability is defined as the proba not finding an address in the cache. For a giv it is simply the ratio of the number of fau total number of references in the trace. The miss probability, the better the replac rithm.

The miss probabilities for various cache s four replacement algorithms are presented 4. From the figure, we see that for small cach

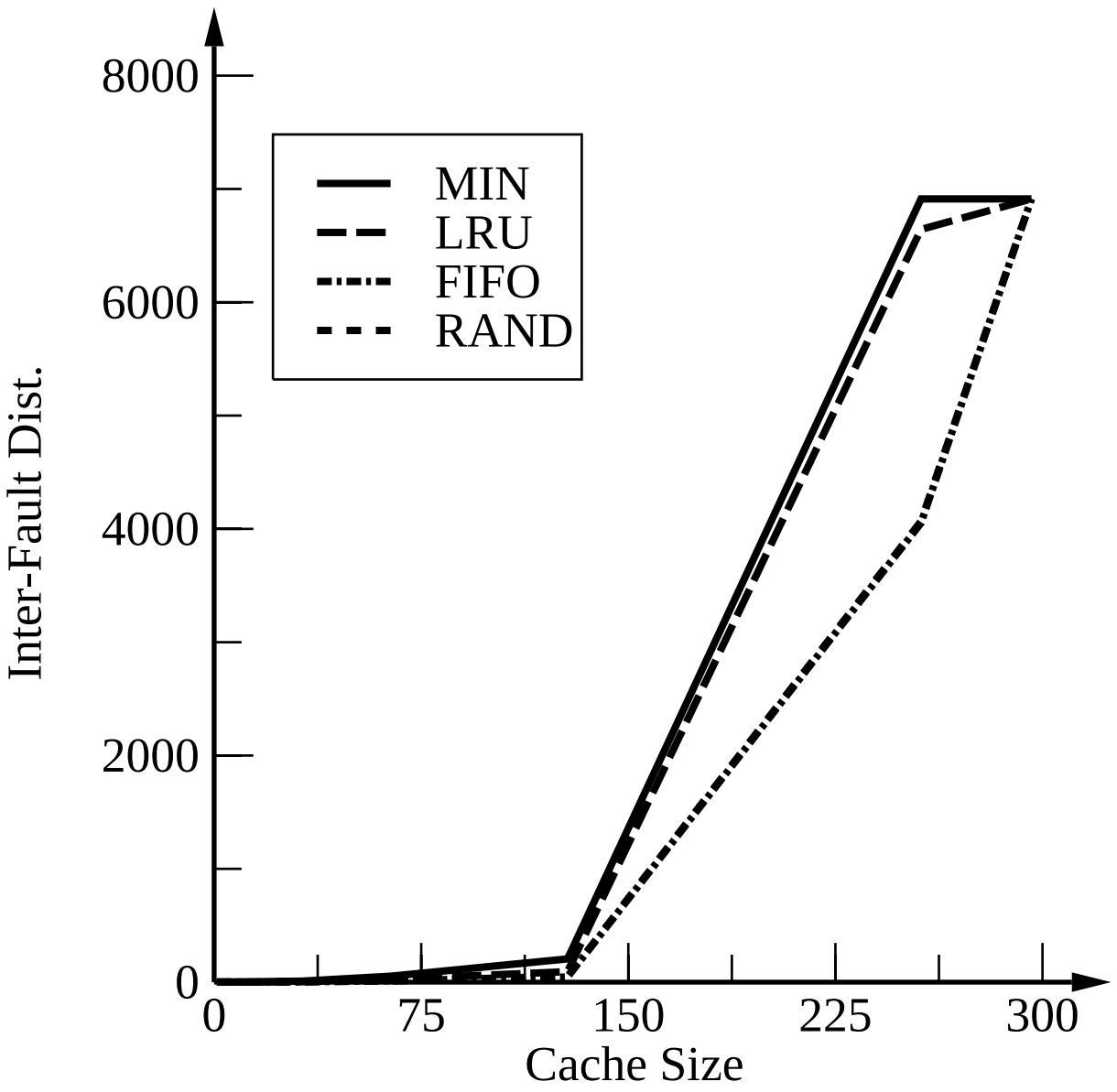

Figure 5: Interfault distances for various cache replacement algorithms.

FIFO, and RAND are not very different for this trace. The miss probability for MIN is better by approximately a factor of two. Thus, there is sufficient room for improvement by designing another replacement algorithm.

For large cache sizes, the miss probability curves are too close to make any inferences. The interfault distance curves provide better discrimination for such sizes.

### 7.2 Interfault Distance

The interfault distance is defined as the number of references between successive cache misses. For a given trace, the average interfault distance can be computed by dividing the total number of references by the number of faults. Thus, average interfault distance is the reciprocal of the miss probability.

Average interfault distances for our trace using the four replacement algorithms are shown in Figure 5.

From the figure we see that for large caches, LRU is close to optimal. FIFO and RAND are equally bad for this trace. Thus, unless one discovers a better replacement algorithm, we can use large caches with the LRU replacement algorithm.

This leads us to wonder what is the optima
size. If a cache is too small, we have a high
If the cache is too large, we do not gain muc
the miss rate is small since we have to searc
a large table. The question of optimal cac
answered by our third metric, normalized se
discussed below.

### 7.3 Normalized Search Time

Caches are useful for several reasons. Firs
have a faster access time then the main da
This is particularly true if the main data
motely located and the cache is local. Sec
may have a faster access method. For exa
caches may be implemented using associate
ories (CAMs). Third, the references have
property so that entries in the cache are m
to be referenced than other entries.

We need to separate the effect of locality an
if there is sufficient locality in the addre
patterns to warrant the use of caches. If
enough locality, one would want to use a ca
if the access time to cache was same as tha
main database, and if the cache used the sam
method (say binary search) that would be u
the main database.

Assuming that the access time and the access
for the cache are the same, we can compute
erage access time with and without cache
the ratio of the two as the metric of contri
performance due to locality alone.

Assuming that a full database of $n$ entrie
generally require a search time proportio
$\log_2(n)$, we have:

$$\text{Time to search without cache} = 1 + \log_2(n)$$

With a cache, if $p$ is the miss probability, w
search through both the cache and and the fu
with probability $p$, and the normalized sear
defined as the ratio:

$$\text{Normalized Search Time} = \frac{\text{Search time with cache}}{\text{Search time without cach}}$$
$$= \frac{(1-p)[1+\log_2(c)] + p[1+\log_2(c)+1+\log_2(n)]}{1+\log_2(n)}$$

The normalized search time for the four rep
algorithms considered is shown in Figure
the figure, we see that with a cache using t replacement algorithm, we could achieve up to 33% less search time than that without caching. The payoff with other replacement algorithms is much less. It is more important to observe, however, that with LRU, FIFO, and RAND, the total search time may be *more* with small caches than that without a cache. For example, with a cache size of 8, these three algorithms would require 20% more search time than without a cache. This trace, therefore, shows a reference pattern in which *caching can be harmful*.

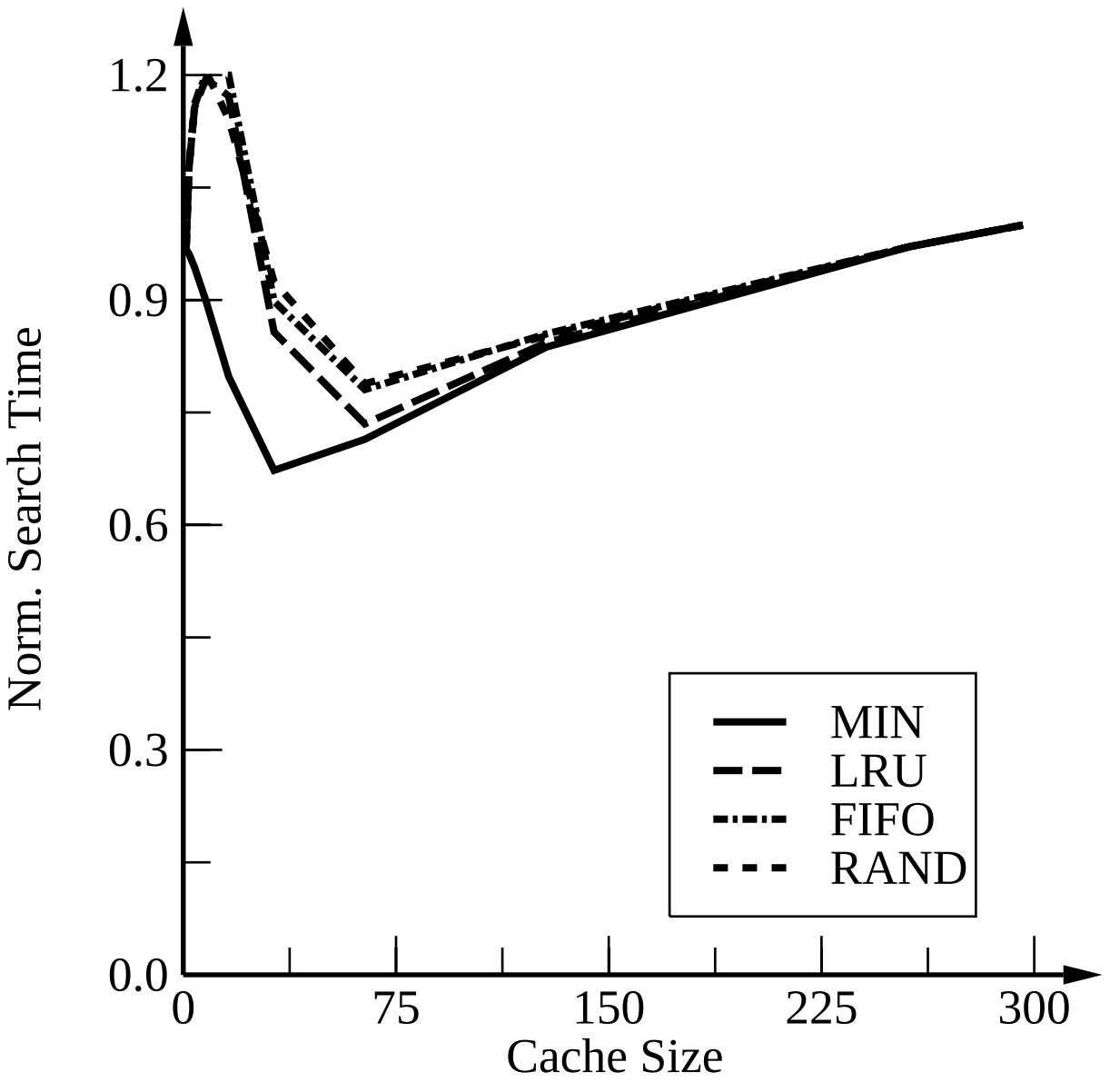

Figure 6: Normalized search time for various cache replacement algorithms.

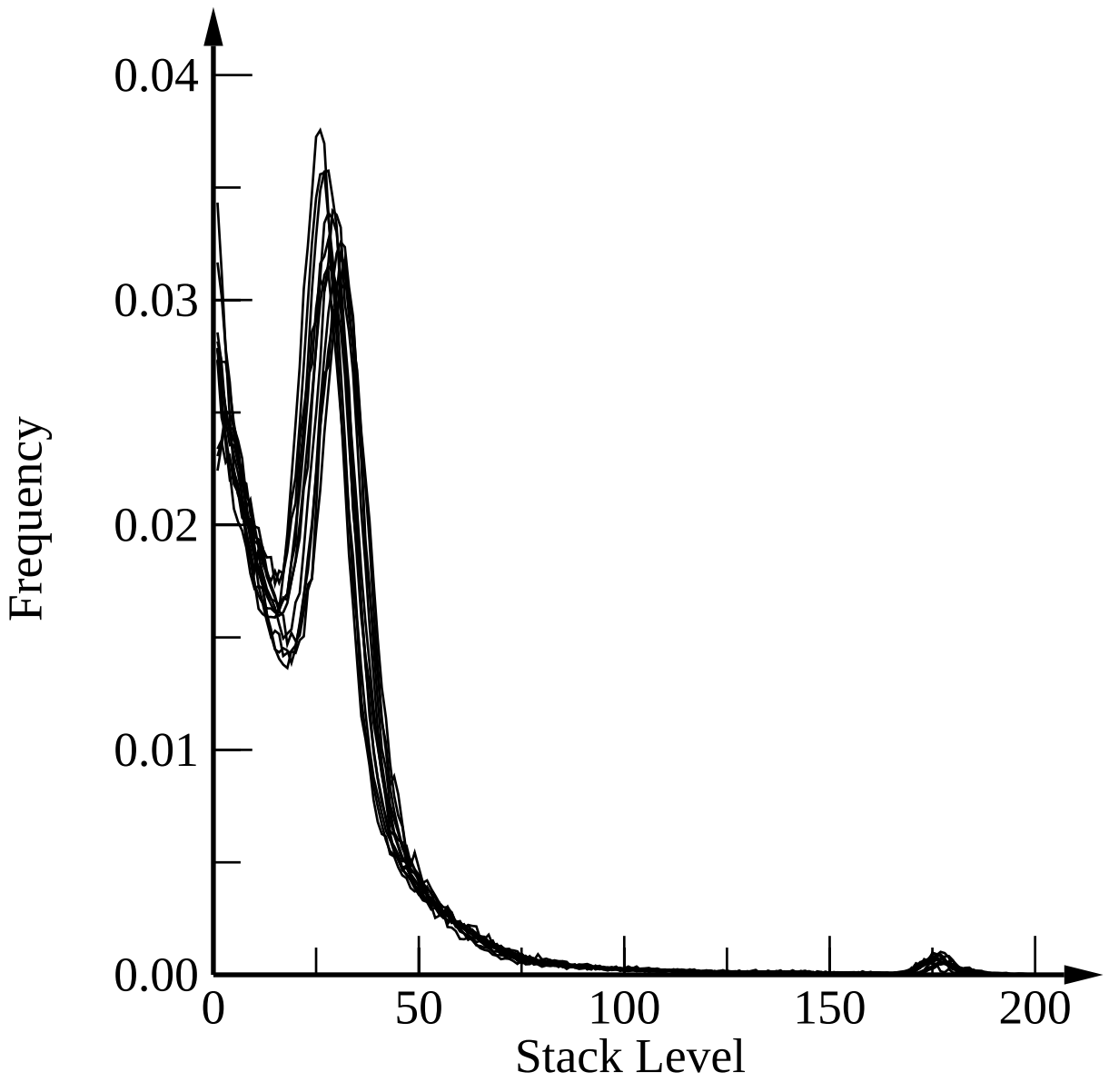

Figure 7: Stack reference frequency.

With a very large cache, the cache does reduce the search time, but the gain decreases as the cache size increases. The optimal cache size for this trace is approximately 64, which produces 20 to 25% reduction in search time.

Earlier measurements at the Massachusetts Institute of Technology [4,9] on a token ring had shown that even a small cache size would provide a big payoff. Therefore, we need to understand what behavior in our environment leads to this different conclusion. We suspect several possibilities. First, the traffic level at M.I.T. is only one tenth of that in our environment. At M.I.T., the traffic level was two million frames per day while in our environment we have that much traffic in one hour. The M.I.T. ring uses an 8-bit address field leading to a maximum of 256 possible addresses on the ring. Actually, there are less than 40 stations on the ring. Our env uses a 48-bit address field and there are 1 tions on the extended LAN. M.I.T. frames ar shorter too. The maximum frame size seen o ring is 576 octets (although the ring allows frames), while the maximum frame size on Eth 1518 octets. A user message is broken into cessive frames resulting in higher persis M.I.T. data. Increased traffic level, more and larger packets could certainly make sma less effective. However, looking at the sta probability density function provided an which we discuss next.

## 8 Stack Reference Frequency

Earlier in Section 6, we showed the cumulat ability distribution function using a stac Instead of adding the probability for succe positions, we plot the probability for indi positions, we get the probability densit (pdf) curve as shown in Figure 7. In this figu have plotted the stack pdf for the complete well as for the 11 subtraces. In all cases, the pdf is not a continuously decreasing fu stead, there is a *hump* around stack position *this environment, the most likely stack position to be referenced is the 30th position and not the stack top.*

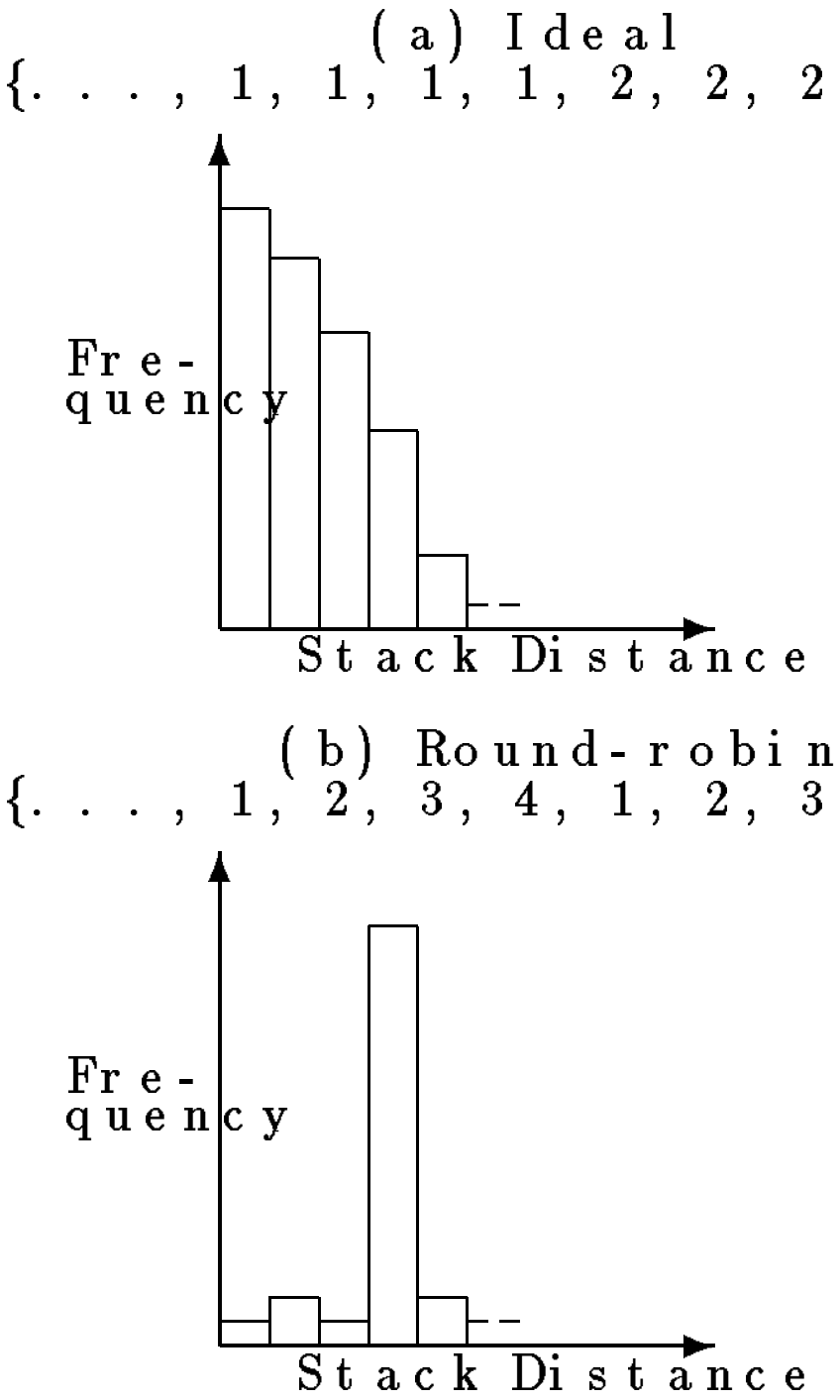

Figure 8: A round-robin reference pattern results in a hump in the stack reference frequency.

LRU is not the best replacement strategy for such a reference string. In general, it is better to replace the address least likely to be referenced again, i.e., the address with minimum probability. For the stack reference probabilities shown in Figure 7, the minimum probability does not always occur at the highest possible stack distance. For example, if the cache size is 30, the address at stack position 15 has a lower probability of reference than that at position 30 and is, therefore, a better candidate for replacement.

One possible cause of the hump could be a round-robin behavior in our reference pattern. To understand this consider two hypothetical reference patterns shown in Figure 8. The first pattern shows a high persistence. Once an address is referenced, it is referenced again several times. Such a reference string would result in a continuously decreasing stack pdf of the type shown in Figure 8a. The second pattern shows a round-robin reference string consisting of $k$ addresses, for instance, repeated over and over again {1, 2, 3, . . . , $k$, 1, 2, 3, . . . , $k$, 1, . . .}. The stack pdf for this string would be an impulse (or Dirac delta) function at $k$, that is, all references would be to stack position $k$.

A mixture of round-robin and persistent tra result in a curve with a hump similar to the served in Figure 7. This round-robin behav be caused by the periodic nature of some of tocols used on our network. In particular, active terminal traffic, which constitutes frames in our trace, uses a protocol calle cal Area Terminals (LAT) [15]. Each LAT se connected to a number of terminals and pro virtual connection to several hosts on the LAN. To avoid sending several small frames, minal input is accumulated for 80 millisec all traffic going to one host is sent as a sin This considerably reduces the number of fr improves the performance of the terminal c cation. A large number of LAT servers trans at regular intervals of 80 milliseconds cou be responsible for the round-robin behavio in the reference pattern.

To verify the above hypothesis, we divided into two subtraces: one consisting entirel tive (LAT) frames, and the other remaining active traffic. The stack pdf for these two s are shown in Figures 9 and 10. Notice that t active traffic exhibits a hump, while the no tive traffic does not. Thus, the interactive seem to be responsible for the hump leadin conclusion that, for environments dominate and similar protocols, one would need eith size equal to the number of LAT servers or to a cache prefetch policy that would bring th dress into the cache just before it is refer

The observation that the noninteractive t a continuously decreasing stack pdf is an i one. Since the LAT traffic is limited to a si tended LAN, it does not go through routers are used to connect several extended LANs area networks. The reference pattern seen is expected to be similar to that of the noni traffic, though we have not yet verified this tion. If this is so, it would be interesti caching would pay off for noninteractive tra We, therefore, analyzed the noninteractiv the next section.

## 9 Analysis of the Noninteract Traffic

In this section, we present the graphs for m bility, interfault distance, and normalize

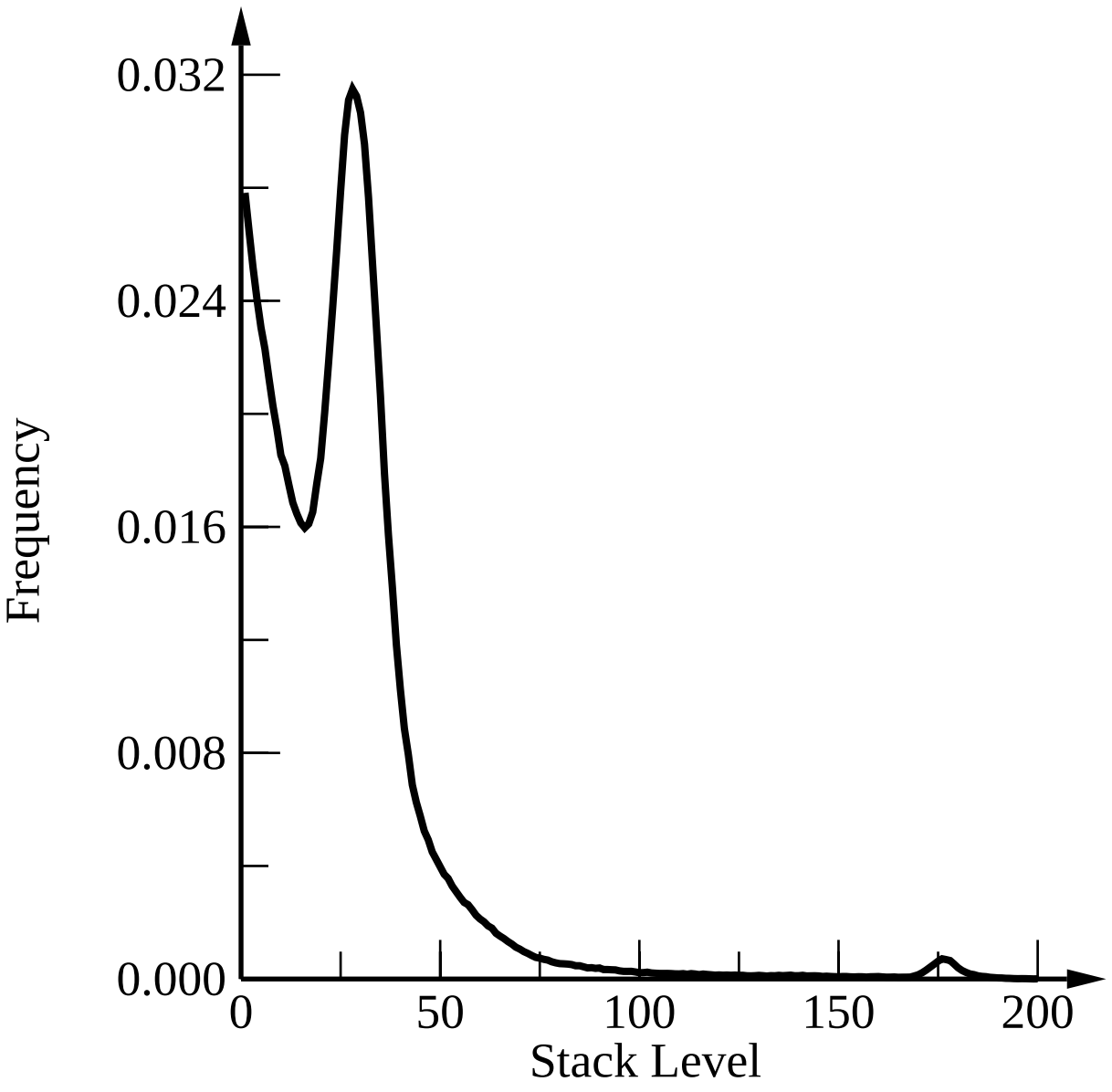

Figure 9: Stack distance density function for LAT traffic.

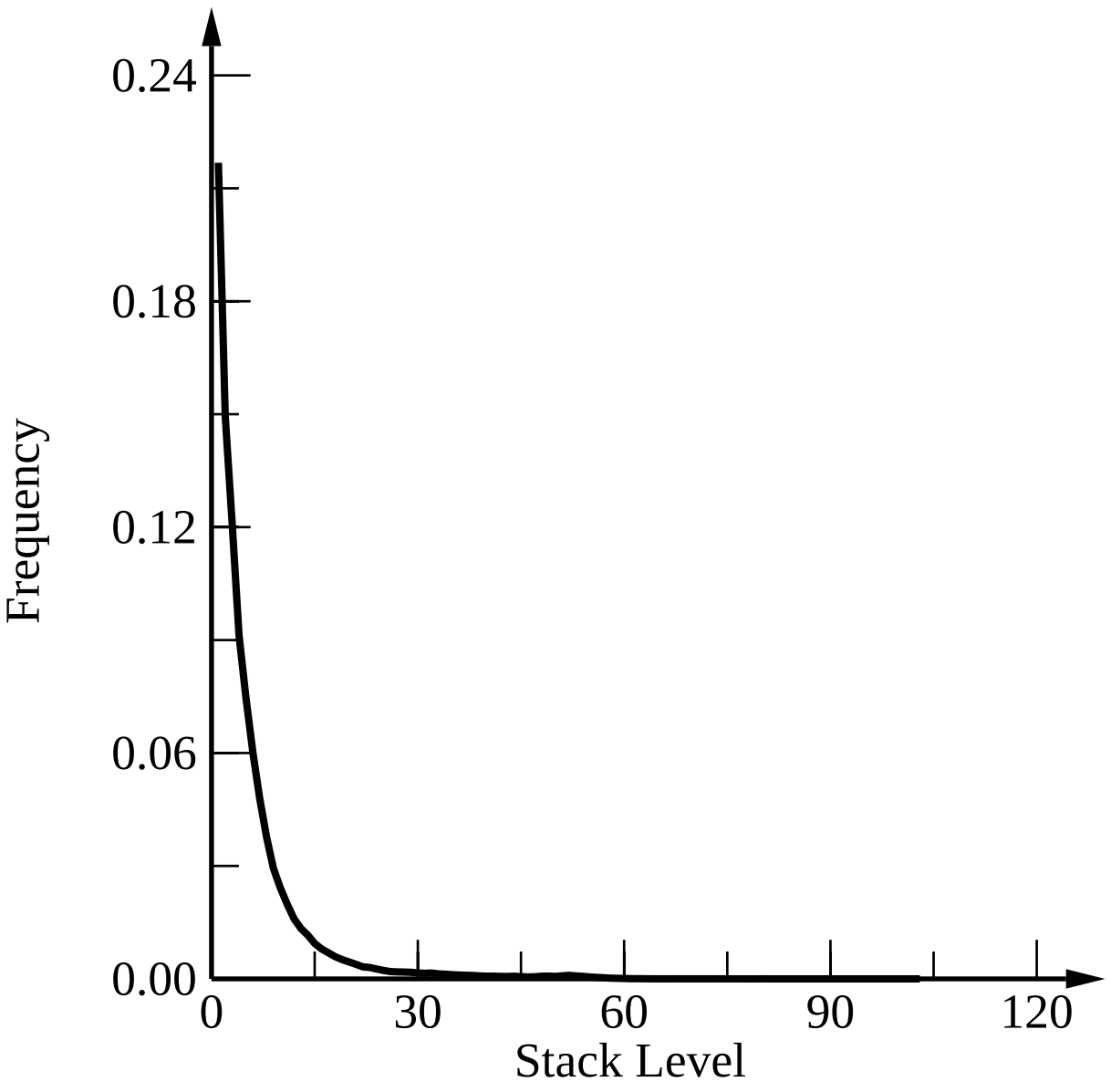

Figure 10: Stack distance density function for noninteractive traffic.

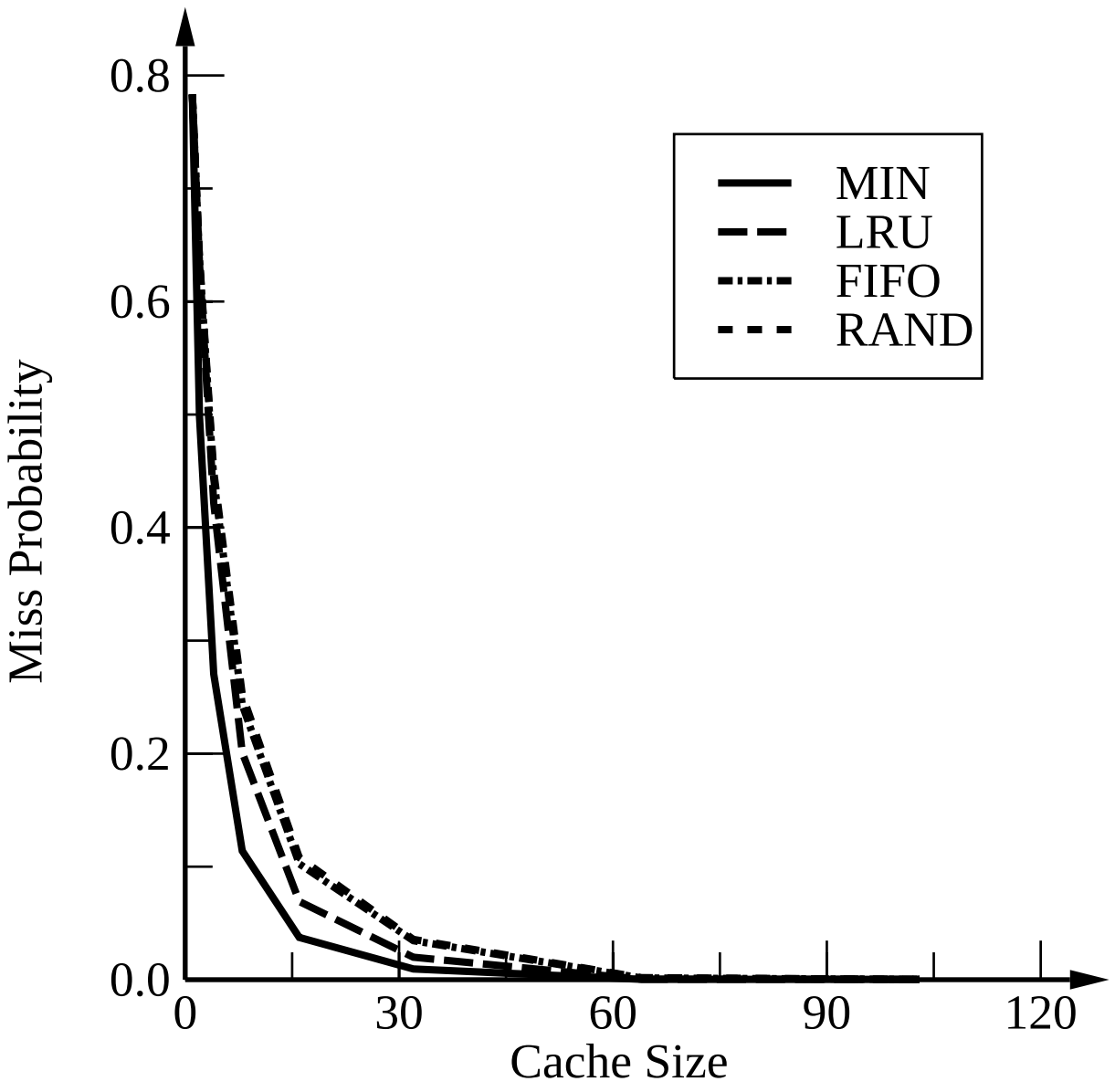

Figure 11: Cache miss probability for noni
frames.

for noninteractive traffic alone. There are t
for repeating the analysis for noninterac
alone. First, as we said earlier, it may gi
indication of behavior of references in ro
ond, it helps us illustrate how some of the c
reached earlier would be different in a diffe
ronment.

Figure 11 shows the miss probability for th
placement algorithms. Notice that even f
caches, LRU is significantly better than F
RAND. This is not surprizing considering
that for any reference trace with nondecrea
pdf, LRU is the optimal cache replacement al
[26]. LRU is optimal in the sense that no ot
tical algorithm can give a lower number of f
any given cache size. MIN does give a lower
of faults and, hence, a lower miss probabili
is due to its knowledge of future references
ence patterns similar to noninteractive tr
fore, we do not need to look for other repl
algorithms. Of course, if LRU is too comple
plement, which is often the case, one woul
simpler algorithms, but that would always c
cost of increased faults.

Figure 12 shows the interfault distances f
replacement algorithms. We see that for la
sizes also, LRU is far superior to FIFO anc

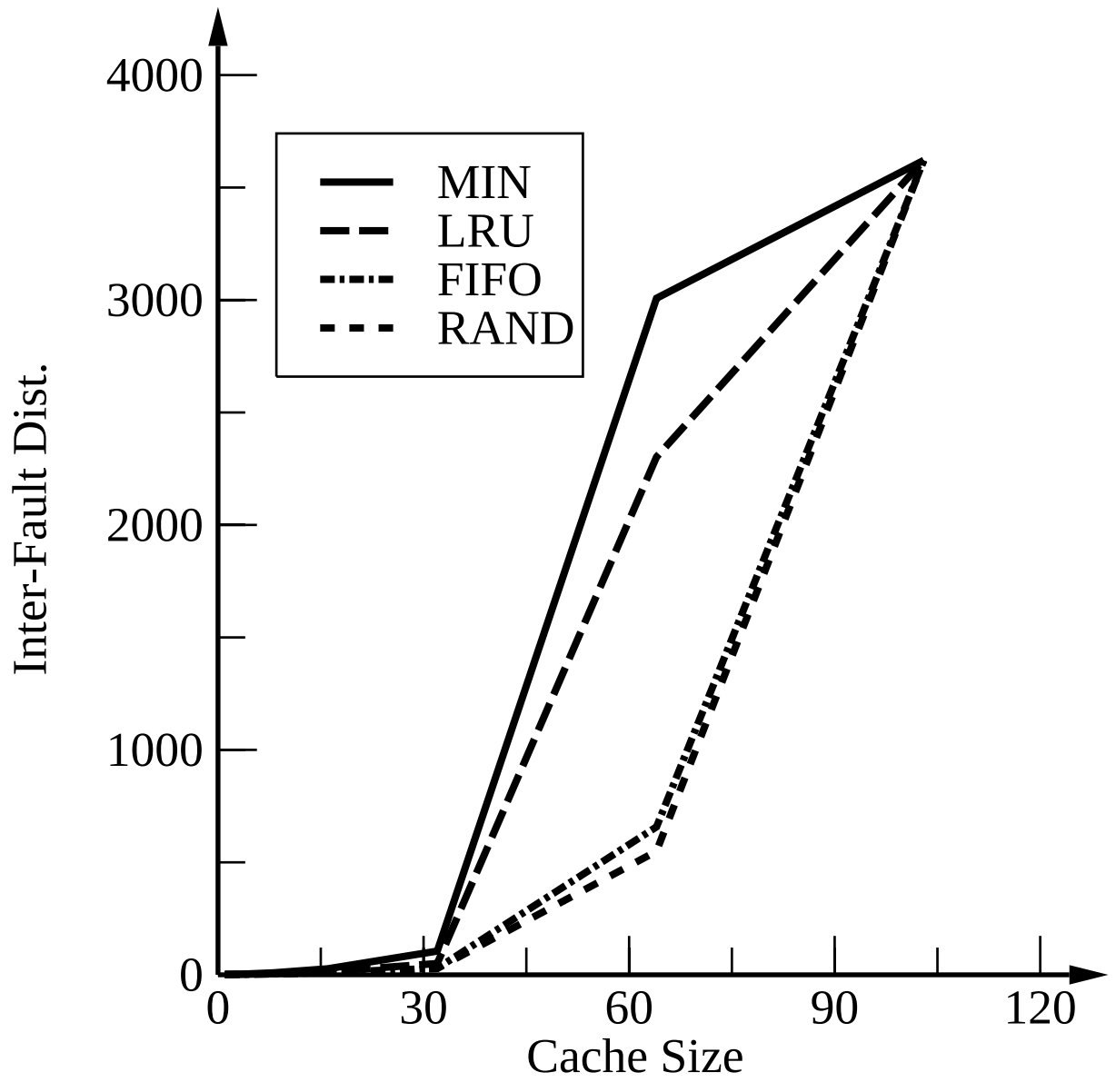

Figure 12: Interfault distances for noninteractive frames.

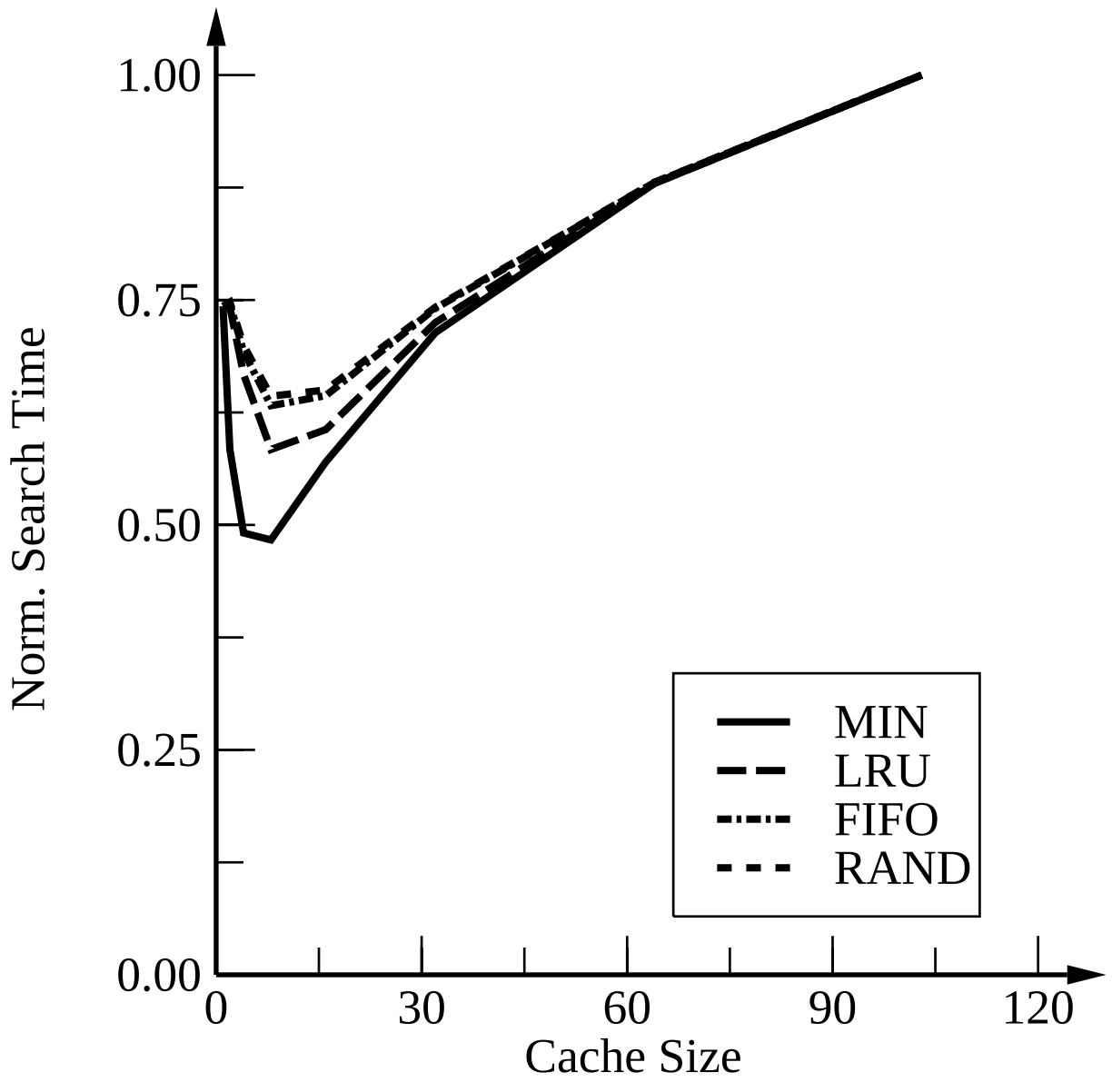

Figure 13: Normalized search time for noni frames.

for this subtrace.

The normalized search time for noninteractive traffic is shown in Figure 13. Notice that for small caches, we now have a valley where we had a peak in Figure 6. Thus, not only are the small caches helpful, they are also optimal. The optimal cache size with LRU is about 8 entries. This reduces the search time by about 40%.

## 10 Other Cache Design Issues

There are many cache design issues that remain to be addressed before caching of network addresses can become a reality. The issues can be classified as cache management, cache structuring, and multicache issues.

Cache management issues relate to algorithms for replacement, fetching, lookup, and deletion. Several replacement algorithms have been compared in this paper. We assumed demand fetching where the address is brought into the cache when it is actually referenced. Prefetching, such as that of source addresses, needs to be analyzed. Address matching strategies, such as the most significant octet first or the least significant octet first may produce different performances. Finally, the issue of deleting addresses periodically needs to be studied.

Processor caches are generally structured Each set consists of several entries. A gi is mapping to a set and the replacement is confined to that set. Two extrem structures are: direct mapped in which eac set consists of only one entry, and fully associat all entries are part of the same set and th mapping.

Another issue related to cache structurin organizing separate caches for different t dresses. For example, in many computer sy instruction and data caches are organized since their reference patterns are so diffe computer networks, one may want to study th of organizing separate caches for group and addresses, separate caches for interactiv teractive traffic, or a separate cache for ea type.

Multicache consistency [12] is also an int sue, particularly in multiport intermediat which each port has a separate cache of addr

Finally, in many networks such as token ring it is important for an intermediate system diately decide whether to set the 'address and 'frame copied' flags in the frame. In su tem, cache lookup time is bounded. It remains to be seen what impact this time bound has on cache management and structuring strategies.

## 11 SUMMARY

As sizes of computer networks grow, we need to find ways to efficiently and quickly recognize destination addresses. Caching is one one such alternative that helps if there is locality in the reference pattern. *Concentration* of references to a small fraction of addresses as well as the *persistence* of the references to recently used addresses help achieve a low miss probability even with small caches.

We reviewed the concepts of spatial and temporal locality along with well-known models such as IRM, working set, and LRU and tried to apply them to destination reference strings.

We compared four different cache replacement algorithms: MIN, FIFO, LRU, and random and discovered that although address traces do have both concentration and persistence, the periodic nature of certain protocols may make the use of small caches ineffective. For those environments where a similar round-robin reference pattern is observed, either we need to develop new cache replacement and fetch algorithms, or to use larger caches.

Some of the observations presented in this paper are limited to our environment and application (bridge caching). However, the methodology is general and can be applied to other environments and problems as well. In particular, it would be interesting to apply it to the study of the reference pattern of the 20-octet addresses used in ISO network layers and the name reference patterns in various name servers and distributed systems.

## 12 Acknowledgments

We would like to thank Ruei-Hsin Hsiao of DEC for helping to gather the trace data. Shawn Routhier, currently of Prime Computers, Inc., also helped in collecting a trace when he was with M.I.T. Although the M.I.T. trace results have not been presented here, the data did help us in getting started on this project. Thanks are also owed to Bill Hawe, Tony Lauck, and other members of Digital's Networking Architecture group for their valuable feedback during the project.

## 13 Appendix: Numerical Results

In this paper, we have presented results graphically wherever possible. To allow easy reading of the values plotted, the same results are now presented in tabular form in this appendix.

Table 5: Average Working Set Size

| Sub- | Window Size | | | | | | |
|---|---|---|---|---|---|---|---|
| trace | 5 | 10 | 20 | 50 | 100 | 200 | 500 |
| 1 | 4.7 | 8.9 | 16.2 | 29.3 | 37.6 | 47.4 | 62.7 |
| 2 | 4.7 | 9.0 | 16.3 | 29.7 | 38.1 | 48.0 | 63.4 |
| 3 | 4.8 | 9.0 | 16.2 | 30.3 | 38.8 | 48.6 | 64.1 |
| 4 | 4.7 | 8.9 | 16.0 | 29.5 | 38.3 | 48.5 | 64.4 |
| 5 | 4.7 | 9.0 | 16.3 | 30.4 | 39.3 | 49.5 | 65.4 |
| 6 | 4.8 | 9.0 | 16.4 | 31.2 | 40.1 | 49.8 | 65.1 |
| 7 | 4.8 | 9.0 | 16.5 | 30.4 | 38.7 | 48.4 | 63.7 |
| 8 | 4.7 | 8.9 | 16.2 | 31.2 | 40.2 | 49.8 | 64.9 |
| 9 | 4.7 | 8.8 | 16.1 | 31.6 | 40.7 | 50.0 | 64.6 |
| 10 | 4.7 | 8.9 | 16.2 | 31.5 | 40.5 | 50.1 | 65.1 |
| 11 | 4.8 | 9.0 | 16.3 | 31.5 | 40.9 | 50.6 | 65.1 |
| Total | 4.7 | 9.0 | 16.3 | 30.5 | 39.3 | 49.0 | 64.4 |

Table 6: Cumulative Frequency for Different Stack Levels

| Sub- | Stack Distance | | | | | | |
|---|---|---|---|---|---|---|---|
| trace | 1 | 2 | 5 | 10 | 20 | 50 | 100 |
| 1 | 0.03 | 0.06 | 0.13 | 0.23 | 0.41 | 0.93 | 0.98 |
| 2 | 0.03 | 0.05 | 0.12 | 0.22 | 0.41 | 0.93 | 0.98 |
| 3 | 0.02 | 0.05 | 0.12 | 0.23 | 0.41 | 0.93 | 0.98 |
| 4 | 0.03 | 0.06 | 0.14 | 0.25 | 0.44 | 0.93 | 0.98 |
| 5 | 0.03 | 0.05 | 0.12 | 0.22 | 0.40 | 0.92 | 0.98 |
| 6 | 0.02 | 0.05 | 0.11 | 0.21 | 0.38 | 0.92 | 0.99 |
| 7 | 0.03 | 0.05 | 0.12 | 0.21 | 0.38 | 0.93 | 0.98 |
| 8 | 0.03 | 0.05 | 0.13 | 0.23 | 0.39 | 0.92 | 0.99 |
| 9 | 0.03 | 0.07 | 0.14 | 0.24 | 0.39 | 0.93 | 0.99 |
| 10 | 0.03 | 0.06 | 0.13 | 0.23 | 0.38 | 0.92 | 0.99 |
| 11 | 0.02 | 0.05 | 0.12 | 0.22 | 0.39 | 0.92 | 0.99 |
| Total | 0.03 | 0.05 | 0.13 | 0.23 | 0.40 | 0.93 | 0.98 |

Table 7: Miss Probability

| Cache Size | MIN | LRU | FIFO | RAND |
|---|---|---|---|---|
| 1 | 0.972 | 0.972 | 0.972 | 0.972 |
| 2 | 0.846 | 0.946 | 0.946 | 0.947 |
| 4 | 0.708 | 0.896 | 0.898 | 0.901 |
| 8 | 0.548 | 0.810 | 0.816 | 0.819 |
| 16 | 0.339 | 0.670 | 0.695 | 0.643 |
| 32 | 0.106 | 0.271 | 0.308 | 0.331 |
| 64 | 0.019 | 0.038 | 0.080 | 0.087 |
| 128 | 0.005 | 0.011 | 0.022 | 0.019 |
| 256 | 0.000 | 0.000 | 0.000 | 0.000 |
| 296 | 0.000 | 0.000 | 0.000 | 0.000 |

Table 8: Average Interfault Distance

| Cache Size | MIN | LRU | FIFO | RAND |
|---|---|---|---|---|
| 1 | 1.0 | 1.0 | 1.0 | 1.0 |
| 2 | 1.2 | 1.1 | 1.1 | 1.1 |
| 4 | 1.4 | 1.1 | 1.1 | 1.1 |
| 8 | 1.8 | 1.2 | 1.2 | 1.2 |
| 16 | 3.0 | 1.5 | 1.4 | 1.6 |
| 32 | 9.5 | 3.7 | 3.2 | 3.0 |
| 64 | 52.8 | 26.4 | 12.5 | 11.5 |
| 128 | 205.3 | 92.8 | 45.4 | 51.6 |
| 256 | 6912.2 | 6642.9 | 4051.5 | 4067.6 |
| 296 | 6912.2 | 6912.2 | 6912.2 | 6912.2 |

Table 9: Normalized Search Time

| Cache Size | MIN | LRU | FIFO | RAND |
|---|---|---|---|---|
| 1 | 0.968 | 0.968 | 0.968 | 0.968 |
| 2 | 0.962 | 1.076 | 1.076 | 1.077 |
| 4 | 0.943 | 1.157 | 1.159 | 1.162 |
| 8 | 0.899 | 1.195 | 1.202 | 1.205 |
| 16 | 0.798 | 1.172 | 1.200 | 1.141 |
| 32 | 0.673 | 0.857 | 0.898 | 0.924 |
| 64 | 0.714 | 0.735 | 0.781 | 0.788 |
| 128 | 0.837 | 0.843 | 0.855 | 0.852 |
| 256 | 0.971 | 0.971 | 0.971 | 0.971 |
| 296 | 1.000 | 1.000 | 1.000 | 1.000 |